\documentclass{nature_mod}
\usepackage{graphicx}
\usepackage{graphics}
\usepackage{amssymb}
\usepackage{hyperref}
\usepackage{xcolor}
\usepackage{amsmath}
\usepackage{newtxtext}
\usepackage[varg,smallerops,upint]{newtxmath}
\usepackage{newfloat}
\usepackage{threeparttablex, tablefootnote}
\usepackage{longtable}
\usepackage{booktabs}

\newcommand{\apjs}{Astrophys. J. Supp.}

\newcommand{\apjl}{Astrophys. J. Let.}
\newcommand{\aap}{Astron. Astrophys.}

\DeclareFloatingEnvironment[fileext=lop]{Extended Data Figure}

\title{Evidence for the Sombrero Galaxy as an Accelerator
of the Highest-Energy Cosmic Rays}

\begin{document}
\maketitle
\author{Hao-Ning He$^{1,2,3}$, Eiji Kido$^{4,1,3}$, Kai-Kai Duan$^{1}$, Yang Yang$^{1}$, Ryo Higuchi$^{3}$, Yi-Zhong Fan$^{1,2}$, Tao Wang$^{5,6}$, Lu-Yao Jiang$^{1,2}$, Rong-Lan Li$^{1,2}$, Ben-Yang Zhu$^{1,2}$, Xiang Li$^{1,2}$, Zi-Qing Xia$^{1}$, Shigehiro Nagataki$^{3,7,8}$, Da-Ming Wei$^{1,2}$, Alexander Kusenko$^{9,10}$}

\begin{spacing}{1.2}
\begin{affiliations}
\item Key Laboratory of Dark Matter and Space Astronomy, Purple Mountain Observatory, Chinese Academy of Sciences, Nanjing 210023, China
\item School of Astronomy and Space Science, University of Science and Technology of China, Hefei 230026, China
\item Astrophysical Big Bang Laboratory (ABBL), Cluster of Pioneering Research (CPR), RIKEN, 2-1 Hirosawa, Wako, Saitama 351-0198, Japan
\item Institute for Cosmic Ray Research, University of Tokyo, Kashiwa, Chiba 277-8582, Japan
\item School of Astronomy and Space Science, Nanjing University, Nanjing, Jiangsu 210093, China
\item Key Laboratory of Modern Astronomy and Astrophysics, Nanjing University, Ministry of Education, Nanjing 210093, China
\item Interdisciplinary Theoretical $\&$ Mathematical Science Program (iTHEMS), RIKEN, Saitama 351-0198, Japan
\item Astrophysical Big Bang Group (ABBG), Okinawa Institute of Science and Technology Graduate University (OIST), 1919-1 Tancha, 
Onna-son, Kunigami-gun, Okinawa 904-0495, Japan
\item Department of Physics and Astronomy, University of California, Los Angeles, CA 90095-1547, USA
\item Kavli Institute for the Physics and Mathematics of the Universe (WPI), The University of Tokyo Institutes for Advanced Study,
The University of Tokyo, Chiba 277-8583, Japan
\end{affiliations}

\begin{abstract}
Ultrahigh-energy cosmic rays (UHECRs) are the highest energy messenger from space, with energies exceeding
1 EeV. Although UHECRs were discovered over 60 years ago, their origin still remains a
mystery.
Pinpointing sources of UHECRs is crucial for understanding the extreme astrophysical processes that accelerate particles to such extraordinary energies. 
We searched for UHECR multiplets via analyzing 17 years of data with energies greater than 40 EeV from the Pierre Auger Observatory.
A spatial association is found between a multiplet of $25.7^{+6.2}_{-7.0}$ cosmic rays and the Sombrero galaxy with a local (global) significance of $4.5~\sigma~(3.3~\sigma)$.
The Sombrero galaxy hosts a supermassive central black hole with a mass of $\sim1\times 10^9 M_{\odot}$ and exhibits large-scale radio lobes and jets. 
Our finding provides critical evidence on active supermassive black holes as the source of the highest-energy cosmic rays.
\end{abstract}

UHECRs are energetic particles from space with energies from 1 EeV to beyond 100 EeV, which are orders of magnitude greater than the energy achieved by human-made particle accelerators. How UHECRs are produced poses serious challenges to current understanding of astrophysics, and holds the possibility to unveil new physics \cite{Sigl2001}. 
Lots of efforts have been devoted to address the origin of UHECRs since their discovery. The Pierre Auger Observatory (PAO) and the Telescope Array (TA) are the two largest UHECR detectors worldwide,
which can measure the spectrum and arrival directions of UHECRs\cite{Auger:2007Sci,TA:2014,Auger:2017Sci,Auger:2020prl,Auger:2022,TA:2024}. 
However, 
the charged particles are deflected by the magnetic fields in the Milky Way and in the intergalactic space, 
causing the observed UHECR directions to deviate from their original source directions, complicating the task of pinpointing the exact sources of UHECRs.
The deflection angles are proportional to the atomic number $Z$, and inversely proportional to the energies of cosmic rays \cite{Golup:2009, Harari:2001, Tinyakov:2005}.   

As a result, the spatial distribution of UHECRs from a single source follows a pattern.
UHECRs with higher rigidity would be more concentrated and located closer to the source, while those with lower rigidity would be more sparsely distributed and located farther away from the source. 
Based on this feature, several methods have been developed to search for such UHECR multiplet patterns by analyzing the arrival directions of UHECRs\cite{Auger:2012multiplets, He:2016, Auger:2015multiplets,Auger:2020,TA:2020}.

In this work, we build a method based on the maximum likelihood approach originally proposed by He et al. \cite{He:2016} to search for UHECR multiplets and constrain the sources by analyzing the arrival directions of the observed UHECRs. 
This method does not rely on specific magnetic field models, which can be highly uncertain. Instead, the deflections caused by the magnetic fields are parameterized and can be constrained simultaneously with the source position.
In this work, we write a new likelihood function of six parameters including the source coordinates ($\alpha_{\rm s}, \delta_{\rm s}$), three parameters characterizing magnetic field deflections ($A_{\rm reg}, A_{\rm ran}, \phi_{\rm reg}$), and the count number of UHECRs in the multiplet ($n_{\rm s}$), as detailed in Methods.
The refined methodology allows searching for the most probable UHECR multiplet from the global data and constraining its source, by taking into account the background contributed by the isotropically distributed UHECRs,
while the original method in He et al.(2016)\cite{He:2016} can only be applied to constrain the source by analyzing the local excess data, and does not take into account the background contribution.

We apply this refined method on the analysis of the PAO {\it Phase I} data (from 1 January 2004 to 31 December 2020) with energies above 40 EeV (1387 events)\cite{Auger:2022}. This energy threshold is adopted on the basis of the discovery of the intermediate-scale anisotropy
at 4 $\sigma$ level from the PAO {\it Phase I} data with energies exceeding 40 EeV reported by the PAO collaboration\cite{Auger:2022}. 
We assume a pure composition for UHECRs above 40 EeV in the following calculation, since the analysis of the mass composition of the PAO events reveals a trend towards less mixed compositions at the highest energies\cite{Mayotte2023}. 

First, we search for the most probable multiplet by using the 6-parameter maximum-likelihood estimation, adopting the reconstructed energies without considering the energy errors.
A UHECR multiplet is successfully identified
with the best-fit cosmic ray count number of $n_{\rm s}\sim25.2$.
Adopting the likelihood-ratio testing, which compares the hypothesis of a multiplet plus an isotropic background versus a background-only null hypothesis (details are described in Methods), we derive the maximum value of the test statistic (TS) for the multiplet as ${\rm TS}_{\rm best-fit}\simeq 41.1$. The best-fit parameters and the maximum TS value are listed in Table \ref{tab:sources}.

To evaluate the significance of the tested hypothese, we simulate datasets of background-only UHECRs, ensuring that their spatial distribution follows the PAO exposure and
their energy distribution follows Gaussian distributions centered on the reconstructed energies of the experimental data with standard deviations that reflect both statistical and systematic errors.
The P values are estimated as the fraction of the background-only simulations with a TS greater than or equal to the maximum TS of the experimental data.
Our analysis reveals that 12 sets of simulations from $5\times10^4$ show a UHECR multiplet with a TS greater than or equal to the value of $\rm TS_{\rm best-fit}$ obtained from observation. This resulted in a P-value of $2.4\times 10^{-4}$, corresponding to a significance level of approximately 3.7 $\sigma$.

\begin{figure*} 
	\centering	\includegraphics[width=0.95\textwidth]{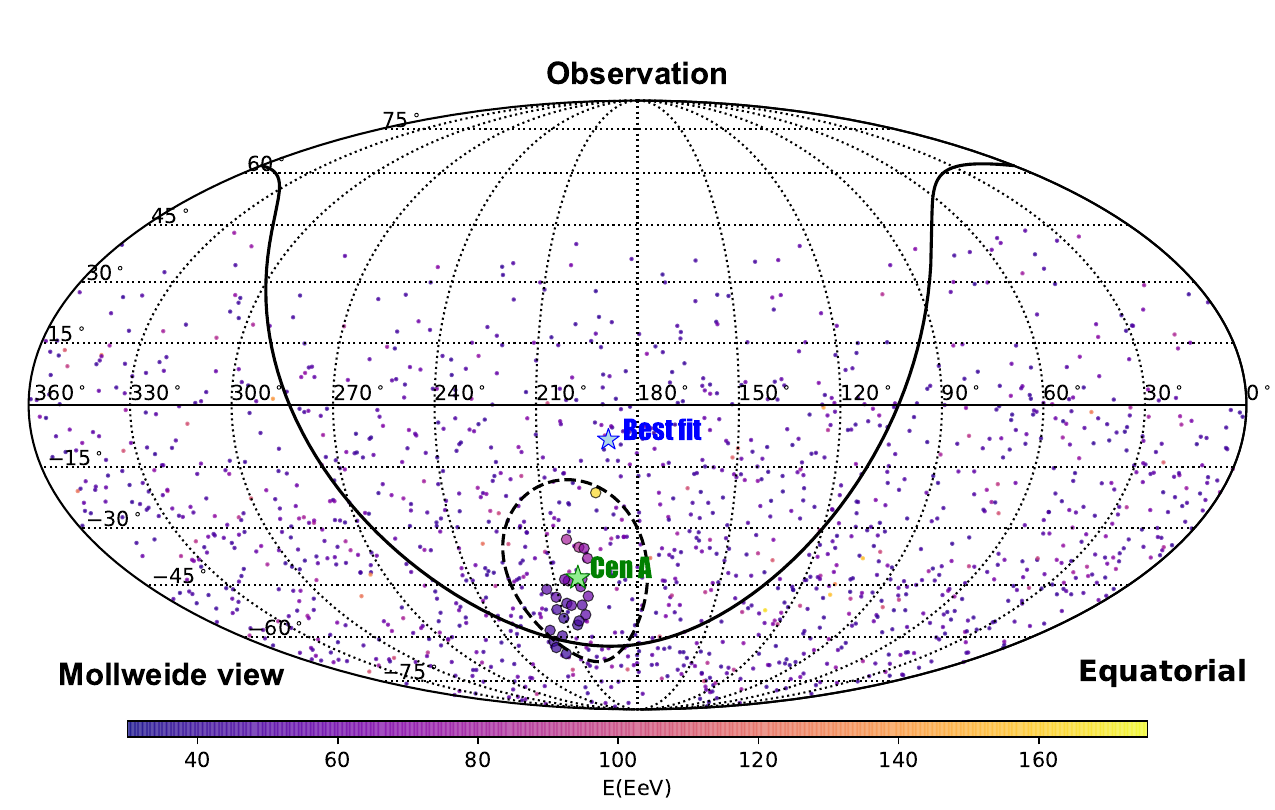} 
	\caption{\textbf{The skymap of the observed UHECR multiplet from the PAO data.}
		The top 25 cosmic rays (filled circles) with energies greater than 40 EeV from the PAO {\it Phase I} data exhibiting the highest probability of originating from the same source at the best-fit position, as indicated by the blue star. The small dots represent the other UHECR events with energy larger than 40 EeV from the PAO {\it Phase I} data. The green star denotes Centaurus A. The black dashed circle denotes the $25^\circ$ radius excess with Centaurus A as the center. The black solid line denotes the galactic plane. The color code denotes the reconstructed energies of UHECRs.}
    \label{fig:fsrc}
\end{figure*}

Adopting the six best-fit parameters listed in Table \ref{tab:sources},
we can identify the top 25 events with the highest likelihood of being from the best-fit source direction, which collectively constitute the UHECR multiplet.
Figure \ref{fig:fsrc} illustrates the distribution of the UHECR multiplet on the sky map.
The multiplet is located in the vicinity of the top-hat excess reported by the PAO collaboration\cite{Auger:2022}, but its spatial distribution differs from the latter.
The observed UHECR multiplet in Figure \ref{fig:fsrc} exhibits a pattern consistent with the theoretical expected distribution of UHECRs with a pure composition originating from a single source after being deflected by the magnetic field. That is,  
the higher-energy UHECRs in the multiplet are closer to the source direction, while the lower-energy ones are farther away from the source direction.
The highest-energy event in the UHECR multiplet, positioned closest to the source, is represented by a yellow filled circle. 
This event, with an energy of $\sim165\pm13$ EeV, is one of the two events with the highest energies reported 
by the PAO collaboration\cite{Auger:2023catalog}.

\subsubsection*{Pinpointing the Sombrero Galaxy as the source}

At the highest energy end, cosmic rays lose energy and mass as a result of interactions with the cosmic microwave background radiation (CMB) and extragalactic background light (EBL) photons as they propagate through extragalactic space, limiting the maximum distance of their sources to a horizon distance \cite{G1966,ZK1966}. 
The horizon distance of their sources, defined here as the distance within which less than 95$\%$ of the number of particles is lost during propagation\cite{Globus2023},
depends on the magnetic fields, the energies, and compositions of the initial injected particles and the observed particles. 
According to the simulations in Globus et al. (2023)\cite{Globus2023}, if the initial injected UHECRs are composed of iron nuclei and the observed mass number $A_{\rm obs}\geq12$,
the horizon distance is estimated to be 36.6 Mpc to observe particles with energy $E_{\rm obs}\geq 150{\rm EeV}$. 
If the initial UHECRs are composed of nitrogen nuclei, the horizon distance is 1.2 Mpc (27.9 Mpc) for $A_{\rm obs}\geq12(A_{\rm obs}\geq1)$.
The largest horizon distance is estimated to be 39.6 Mpc if the initial UHECRs are composed of protons, in which case the composition of the observed UHECRs is also proton.
Here we adopt the most conservative horizon distance 39.6 Mpc as the maximum distance of the source of the $\sim 165\pm 13$ EeV event.

\begin{figure*} 
	\centering
	\includegraphics[width=0.95\textwidth]{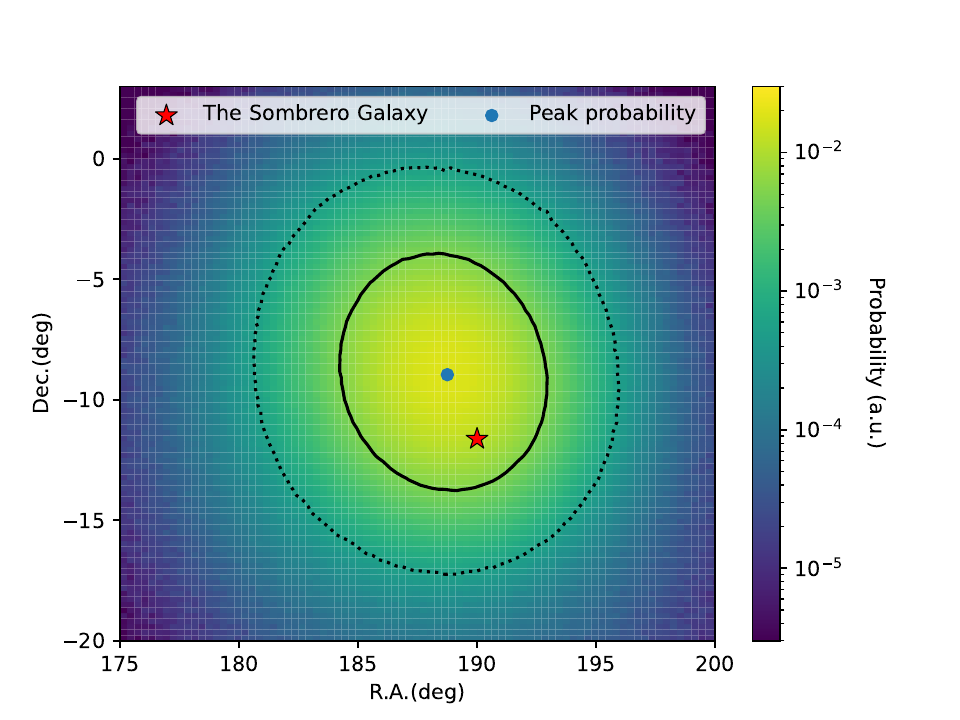} 
	\caption{\textbf{The confidence regions of the source coordinates.}
		The $68\%$ and $95\%$ confidence regions of the source coordinates, taking into account the errors of the reconstructed energies of the UHECRs, are indicated by solid and dotted lines, respectively. The red star denotes the coordinates of the Sombrero Galaxy. The color code denotes the probability in arbitrary units. The blue dot denotes the coordinates with the peak probability. }
    \label{fig:4para}
\end{figure*}

Potential accelerators for such high-energy cosmic rays include large-scale accretion shocks surrounding galaxy clusters, superwinds emanating from starburst galaxies, and jets/outflows/lobes associated with active black holes\cite{Anchordoqui:2018qom}. 
Accordingly, we seek to identify potential candidate sources situated within the 95$\%$ confidence level region of the constrained source coordinates, within the horizon distance of 39.6 Mpc, from catalogs of X-ray selected galaxy clusters \cite{Koulouridis2021}, starburst galaxies\cite{Auger:2022},
hard X-ray active galactic nuclei (AGNs) observed by {\it Swift}/BAT \cite{Baumgartner2013,Auger:2022}, gamma-ray AGNs observed by {\it Fermi}/LAT \cite{Ajello2017}, and nearby galaxies with jets or lobes \cite{vanVelzen2012,Karachentsev2013}.
The 95$\%$ confidence level region of the source coordinates is calculated taking into account the systematic and statistical errors of the reconstructed energy (details are described in Methods).
We find that in the 95$\%$ confidence region of the source coordinates, the Sombrero Galaxy (M 104, NGC 4594) is the only candidate within 39.6 Mpc from the catalogs mentioned above.
The Sombrero Galaxy is located within the 68$\%$ confidence region, as illustrated in Figure \ref{fig:4para}.

By setting the Sombrero Galaxy as the source, the maximum value of the TS of a UHECR multiplet associated with the Sombrero Galaxy is obtained as $\rm TS_{\rm M104}=37.6$, where the reconstructed energies of the observed UHECRs are adopted.
The analysis constrains the deflection effect of the magnetic fields
on the UHECRs originating from the Sombrero Galaxy. The constrained parameters are listed in Table \ref{tab:sources}.
Fixing the source to be the Sombrero Galaxy, we derived the local P value as $7\times10^{-6}$ (there are 7 out of $10^6$ background-only simulation data sets with a TS greater than or equal to $\rm TS_{\rm M104}$), which corresponds to the local significance of $4.5~\sigma$.
To calculate the global significance considering the spatial look-elsewhere effect, we apply the 6-parameter fitting on the simulation data sets, where the source coordinates are set as free parameters.
There are 49 out of $5\times10^4$ simulation data sets with a TS greater than or equal to $\rm TS_{\rm M104}$, which corresponds to a post-trial P value of $9.8\times10^{-4}$ and a global significance of $3.3~\sigma$.

\subsubsection*{ Simulating a UHECR multiplet from the Sombrero Galaxy}
\begin{figure*} 
	\centering
 \includegraphics[width=0.95\textwidth]{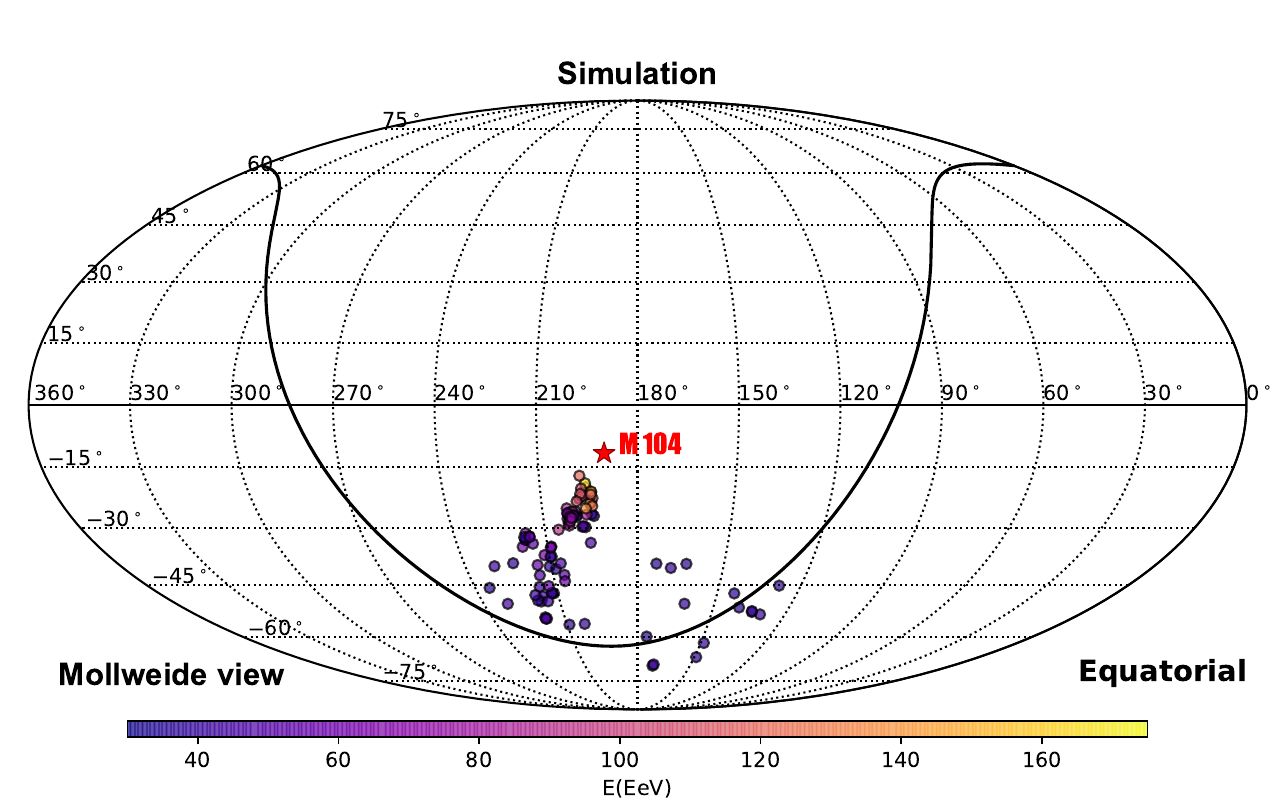}
	\caption{\textbf{The skymap of the simulated UHECR multiplet from the Sombrero Galaxy.}
		The simulated distribution of a multiplet of 100 UHECRs from the Sombrero galaxy assuming the atomic number as Z=7, predicted via the lensing function in CRpropa\cite{Batista2022}, taking into account the Jasson $\&$ Farrar (2012)'s GMF model\cite{Jansson2012a, Jansson2012b} and the PAO's exposure\cite{Auger:2022}. The initial number distribution of UHECRs in the UHECR multiplet is assumed to follow $dN/dE\propto E^{-2}$. The red star denotes the Sombrero Galaxy. The black solid line denotes the galactic plane. The color code denotes the energies of UHECRs.}
    \label{fig:lensmap}
\end{figure*}

To further verify that the observed pattern of the UHECR multiplet illustrated in Figure \ref{fig:fsrc} can be formed in reality, we simulate UHECR multiplets from the Sombrero Galaxy being deflected by a realistic Galactic Magnetic Field (GMF) model \cite{Jansson2012a,Jansson2012b},
by adopting the lensing function of the public software CRpropa \cite{Batista2022}.
Figure \ref{fig:lensmap} depicts a simulated UHECR multiplet with 100 events from the Sombrero galaxy assuming the atomic number $Z=7$ as an illustrative example. 
The particle charge is entangled with the magnetic field strength when accounting for the UHECR deflections. Consequently, a heavier composition coupled with a weaker magnetic field strength could yield an equivalent simulation.

A notable similarity can be seen between the observed UHECR multiplet derived from the experimental data (as illustrated in Figure \ref{fig:fsrc}) and the simulated multiplet (in Figure \ref{fig:lensmap}). 
In particular, the directions of systematic shifts of UHECRs exhibit a high degree of consistency between observation and simulation. 
In addition, the two multiplets exhibit similar spatial distribution characteristics.
Specifically, the higher-energy cosmic rays are clustered in closer proximity to the source direction, whereas the lower-energy ones are situated at a greater distance from the source direction.
The similarity between the observation and the simulation further supports the association between the UHECR multiplet and the Sombrero Galaxy.

\subsubsection*{Astrophysical interpretation}
\begin{figure*} 
	\centering
	\includegraphics[width=0.95\textwidth]{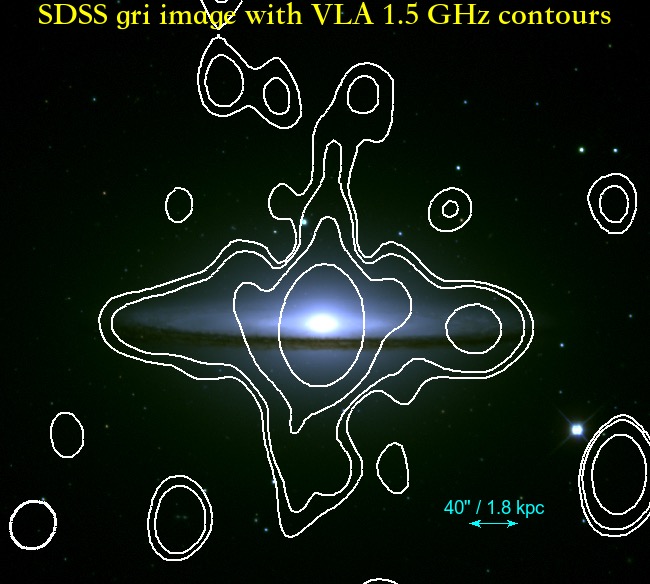} 

	\caption{\textbf{The color image of the Sombrero Galaxy.}
		A color composite image showcasing the global morphology of M 104, constructed by stacking Sloan Digital Sky Survey (SDSS) images in the g, r, and i filters, represented in blue, green and red, respectively. White contours indicate the intensity of the VLA 1.5 GHz D-configuration data\cite{Yang2024}, plotted at the rms error levels of $45~ \rm \mu Jy~ beam^{-1}\times[3,5,15,100]$. }
	\label{fig:M104} 
\end{figure*}

The Sombrero Galaxy, which looks like a Mexican hat,
is located at a distance of $9.5\pm0.13\pm 0.31~\rm Mpc$\cite{McQuinn2016}.
It hosts one of the most massive central black holes in the nearby universe, with a mass of $M_{\rm BH}\sim 1\times10^{9} M_\odot$ \cite{Kormendy1988,Kormendy1996}.
An active supermassive black hole has been proposed to be an extreme particle accelerator, capable of accelerating particles to energies exceeding 100 EeV\cite{Neronov2009,Tursunov2020}.  
Moreover, this galaxy exhibits jets and lobes at different scales ranging from 0.01 parsec to 10 kiloparsec, indicating the activity of the central supermassive black hole\cite{Gallimore2006,Li2011,Hada2013,Mezcua2014,Yang2024,Yan2024}.
The observations indicate
an intrinsic jet velocity of $\lesssim 0.2c$ \cite{Hada2013} and a total jet power of $2.3\times10^{42}~{\rm erg~s^{-1}}$ \cite{Mezcua2014}.
Gallimore et al. \cite{Gallimore2006} identified a kiloparsec-scale radio structure believed to be a supermassive black hole-driven jet or outflow, which is also consistent with the X-ray observation \cite{Li2011}.
Lately, bipolar radio lobes have been detected on the $10$-kiloparsec scale \cite{Yang2024}, as shown in Figure \ref{fig:M104}. 
Based on the assumption of the equipartition between the energy densities
of the CRs and the magnetic fields,
the magnetic field strength in the radio lobes is estimated to be $5-7~\rm\mu G$ and $26-51~\rm\mu G$ for the kiloparsec- and hundred-parsec-scale regions, respectively\cite{Yang2024}.
Simulations carried out by Alves et al.\cite{Alves2018} and Mattews et al.\cite{Mattews2019} suggest that the supermassive blackhole or large-scale jets/lobes can accelerate UHECRs.
The maximum energy of the accelerated cosmic rays is estimated as the Hillas energy \cite{Hillas1984}
\begin{equation}
    E_{\rm max}=234~{\rm EeV}~\left(\frac{Z}{26}\right)\left(\frac{B}{5\mu\rm G}\right)\left(\frac{v_{\rm sh}}{0.2c}\right)\left(\frac{r}{10 \rm kpc}\right),
\end{equation}
\noindent where $B$ is the magnetic field strength at the acceleration site, $v_{\rm sh}$ is the velocity of the shock,
and $r$ is the shock scale. 
According to the Hillas estimation, it is feasible for large-scale jets/lobes powered by the supermassive black hole of the Sombrero Galaxy to accelerate heavy nuclei to energies exceeding 100 EeV.

Adopting the PAO exposure for vertical and inclined events at different declinations, 
the observed mean flux of the UHECR multiplet associated with the Sombrero is estimated to be approximately $1.9\times10^{-12}{\rm erg~ cm^{-2}~ s^{-1}}$. If we do not consider any spatial or temporal enhancement or reduction effects on the UHECR flux due to the magnetic fields, the luminosity of UHECRs with energy exceeding 40 EeV injected by a source at distance $D$ is estimated as $L_{\rm UHECRs}=1.9\times10^{40}{\rm erg ~s^{-1}}\left(\frac{D}{9.5{\rm Mpc}}\right)^{-2}$. Therefore, the required energy of UHECRs for the Sombrero Galaxy is about 1$\%$ of its estimated jet power.

\subsubsection*{Comparison between the Sombrero Galaxy and Centaurus A}
Apart from the Sombrero Galaxy, another nearby AGN known for its jets and large-scale lobes, Centaurus A at a distance of 3.8 Mpc, 
with a substantial jet power of approximately $10^{43}~{\rm erg~s^{-2}}$\cite{Wykes2013},
has been widely discussed as a potential source of UHECRs. 
However, Figure \ref{fig:fsrc} shows that Centaurus A is far from the best-fit source position, indicating that it is not the source of the UHECR multiplet that we observed.
A specific search for the UHECR multiplet associated with Centaurus A has been conducted and the results are presented in Table \ref{tab:sources}. Despite its large jet power of approximately $10^{43}~{\rm erg~s^{-1}}$\cite{Wykes2013}, the maximum TS value for the association of Centaurus A with a UHECR multiplet is found to be $18.2$, which is significantly lower than that for the Sombrero Galaxy.
The difference between the analysis results for the Sombrero Galaxy and Centaurus A may be attributed to the distinct magnetic field deflection effects on UHECRs from these two sources, which are a consequence of their differing positions in the sky. 
The galactic coordinates of Centaurus A and the Sombrero Galaxy are $(l,b)_{\rm CenA}=(309.5^\circ,19.4^\circ)$ and $(l,b)_{\rm M104}=(298.5^\circ,51.1^\circ)$, respectively.
The lower galactic latitude of Centaurus A results in more pronounced random GMF deflections on the UHECRs that originate from it
\cite{Han2017, Xu2024}. 
As a result, the UHECR multiplet from Centaurus A is more scattered, making it more difficult to be identified from the data.
Conversely, the higher galactic latitude of the Sombrero Galaxy results in weaker random GMF deflections on the associated UHECRs\cite{Han2017, Xu2024}, 
making it easier to distinguish the UHECR multiplet from the isotropic background. 
This is indeed consistent with the relatively low value of the random magnetic field deflections on UHECRs from the Sombrero Galaxy, constrained by our data analysis, as listed in Table \ref{tab:sources}.

\subsubsection*{Conclusions and discussions}
In this work, we build a method to search for the UHECR multiplet and to constrain the source coordinates, which provides insight on the origins and acceleration mechanisms of these highest-energy particles in the cosmos.  
We find a UHECR multiplet that includes a 165 EeV from the PAO {\it Phase I} data, and constrain the source location to a small error region. Within the small error region of the source and the conservative horizon distance for the 165 EeV event, we pinpoint the source to be the Sombrero Galaxy 
and constrain the effect of the magnetic field on the trajectories of UHECRs from the source. 
The local significance of the UHECR multiplet associated with the Sombrero Galaxy is $4.5~\sigma$,
and the global significance considering the look-elsewhere effect is $3.3~\sigma$.
This result provides evidence for the acceleration of UHECRs attributed to the activity of the central supermassive black hole.

The method built in this work sets source coordinates and the impacts of magnetic fields on deflections as free parameters, thereby ensuring that the results remain unaffected by uncertainties of the specific magnetic field models. 
The primary uncertainties lie in the reconstructed energies and the compositions of UHECRs.
Future observations of UHECRs will increase the statistics of the data, enhance the energy resolution, and provide more precise composition measurements\cite{GRAND2020, AugerPrime2022}.
This will facilitate a more significant association between the UHECR multiplet and its source.

\section*{Methods}
\subsubsection*{The PAO {\it Phase I} data}
The PAO, located in Argentina, is the largest UHECR detector in the world. The public PAO {\it Phase I} data contains 2635 events with the reconstructed energies exceeding 32 EeV, recorded with the surface detector (SD) array from January 1, 2004 to December 31, 2020. The SD array exhibits full efficiency above 3 EeV (4 EeV) for the vertical (inclined) events with $100\%$ duty cycle.
The vertical events are those with zenith angles $\theta_{\rm z}<60^\circ$, and the inclined events are those with zenith angles $60^\circ\leq\theta_{\rm z}<80^\circ$.
The exposure can be calculated in a geometrical way for vertical events and inclined events, respectively. 
The systematic and statistical errors of the reconstructed energy is about $14\%$ and $7\%$, respectively\cite{Auger:2020}.
In this study, we apply our analysis to data comprising 1,067 vertical events and 320 inclined events with reconstructed energies greater than 40 EeV, along with the corresponding exposure\cite{Auger:2022}.
In our method, a pure composition is assumed for UHECRs greater than 40 EeV, as the analysis of the mass composition of PAO events reveals a trend towards less mixed compositions at higher energies\cite{Mayotte2023}. 

\subsubsection*{The likelihood function of searching for the UHECR multiplet}

Given our limited understanding on the GMF (note that the intergalactic magnetic fields are likely too weak to severely deflect UHECRs \cite{Xia:2022uua}),  
following He et al. (2016) \cite{He:2016}, 
{\rm we do not adopt any detailed GMF models but rather categorize the magnetic fields into a regular component and a random component based on their impact on UHECR trajectories.}
The regular magnetic field component will shift UHECRs with energy $E$ and charge $Z$ to the same direction systematically with deflection angles as
\begin{equation}
   \sigma_{\rm reg}\simeq 5^\circ Z\frac{100\, {\rm EeV}}{E}\frac{D}{10\rm kpc}\frac{B_{{\rm reg},\perp}}{1 \mu\rm G}=A_{\rm reg}\frac{100\, {\rm EeV}}{E}, 
\label{eq:Areg}
\end{equation}
where $B_{{\rm reg},\perp}$ is the strength of the magnetic field perpendicular to the propagation path,
$D$ is the propagation length. 
The random magnetic field component
will cause UHECRs to walk randomly. 
The root mean squared (rms) deflection angles of particles can be written as
\begin{eqnarray}\label{eq_dif}
\sigma_{\rm ran}&\simeq & 0.36^{\circ}Z  \frac{100\, \rm EeV}{E} \left(\frac{D}{10{\rm kpc}}\right)^{\frac{1}{2}}
\left(\frac{D_{\rm c}}{100{\rm pc}}\right)^{\frac{1}{2}}\frac{B_{\rm ran}}{1~{\rm \mu \rm G}}\nonumber\\
&=& A_{\rm ran} \frac{100\, \rm EeV}{E},
\end{eqnarray}
where $B_{\rm rms}$ and $D_{\rm c}$ are the rms strength and the coherence length of the random magnetic field.
When a group of UHECRs with a homogeneous composition emanates from a single point source, the deflection angles they experience are inversely proportional to their energy.
The part of deflections influenced by the magnetic field characteristics and the atomic number $Z$, are described as parameters of $A_{\rm reg}=5^\circ Z({D}/{10\rm kpc})({B_{{\rm reg},\perp}}/{1 \mu\rm G})$ and $A_{\rm ran}=0.36^{\circ}Z  \left({D}/{10{\rm kpc}}\right)^{{1}/{2}}
\left({D_{\rm c}}/{100{\rm pc}}\right)^{{1}/{2}}({B_{\rm ran}}/{1~{\rm \mu \rm G}})$.

\begin{figure*} 
	\centering
	\includegraphics[width=0.6\textwidth]{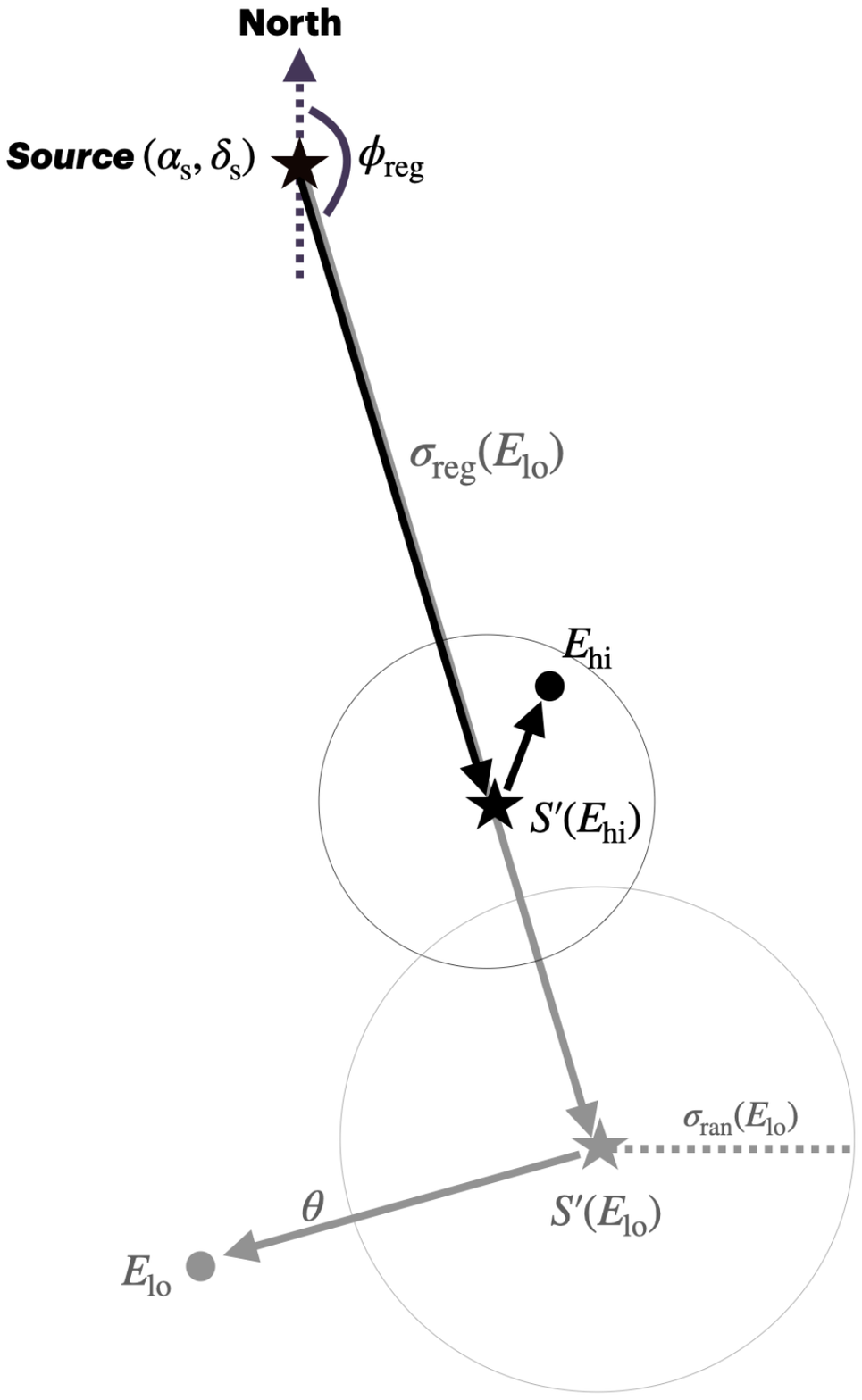} 

	\caption{\textbf{The sketch showing the deflections of UHECRs of different energies.}
		The sketch showing the deflections of two UHECRs with energies of $E_{\rm lo}$ and $E_{\rm hi}$, and coordinates of $(\alpha_{\rm lo},\delta_{\rm lo})$ and $(\alpha_{\rm hi},\delta_{\rm hi})$, respectively, where $E_{\rm lo}<E_{\rm hi}$.}
	\label{fig:cartoon} 
\end{figure*}

We define the coordinates of UHECRs with energy $E$ with a pure composition after being shifted by the regular magnetic field component as the apparent source coordinate $S^\prime(E)$, as shown in Figure \ref{fig:cartoon}.
The deflection of a UHECR with an energy of $E$ originating from a point source
with coordinates of $(\alpha_{\rm s},\delta_{\rm s})$
can be described as a systematic shift from the original source to $S^\prime (E)$, then plus a random walk in a random direction from $S^\prime (E)$.
The clockwise angle between the north direction and the systematic shifted direction is denoted as $\phi_{\rm reg}$.
For the i-th UHECR with energy of $E_i$, 
the apparent source is marked as $S'_i(E_i)$.
The separation angle between the arrival direction of the i-th UHECR ($\alpha_i,\delta_i$) and the apparent source $S^\prime_i(E_i)$ can be measured as $\theta_i(E_i,\alpha_{ i},\delta_{ i}|\alpha_{\rm s},\delta_{\rm s}, A_{\rm reg}, \phi_{\rm reg})$.
The distribution of $\theta_i$ follows a two-dimensional isotropic Gaussian distribution with $\kappa$ as invariance\cite{Fisher1953}, where $\kappa=2/\sqrt{\sigma^2_{\rm ran}+\sigma^2_{\rm ang}}$ with $\sigma_{\rm ang}\simeq 1^\circ$ as the conservative angular resolution\cite{Auger:2022}.
Therefore, the probability of the $i$-th event being from the source can be written as
\begin{equation}
    f_{\rm src}(E_i,\alpha_{ i},\delta_{ i}|\alpha_{\rm s},\delta_{\rm s}, A_{\rm reg},A_{\rm ran}, \phi_{\rm reg})=\frac{\kappa}{4\pi\mathrm{sinh}\kappa}\mathrm{e}^{\kappa\mathrm{cos}\theta_i}.
\label{Eq:fsrc}
\end{equation}
Figure \ref{fig:cartoon} presents the feature that $\sigma_{\rm reg}$ and $\sigma_{\rm ran}$ are larger for the UHECR with a lower energy of $E_{\rm lo}$ than for the UHECR with the higher energy of $E_{\rm hi}$.

\begin{figure*} 
	\centering
	\includegraphics[width=0.48\textwidth]{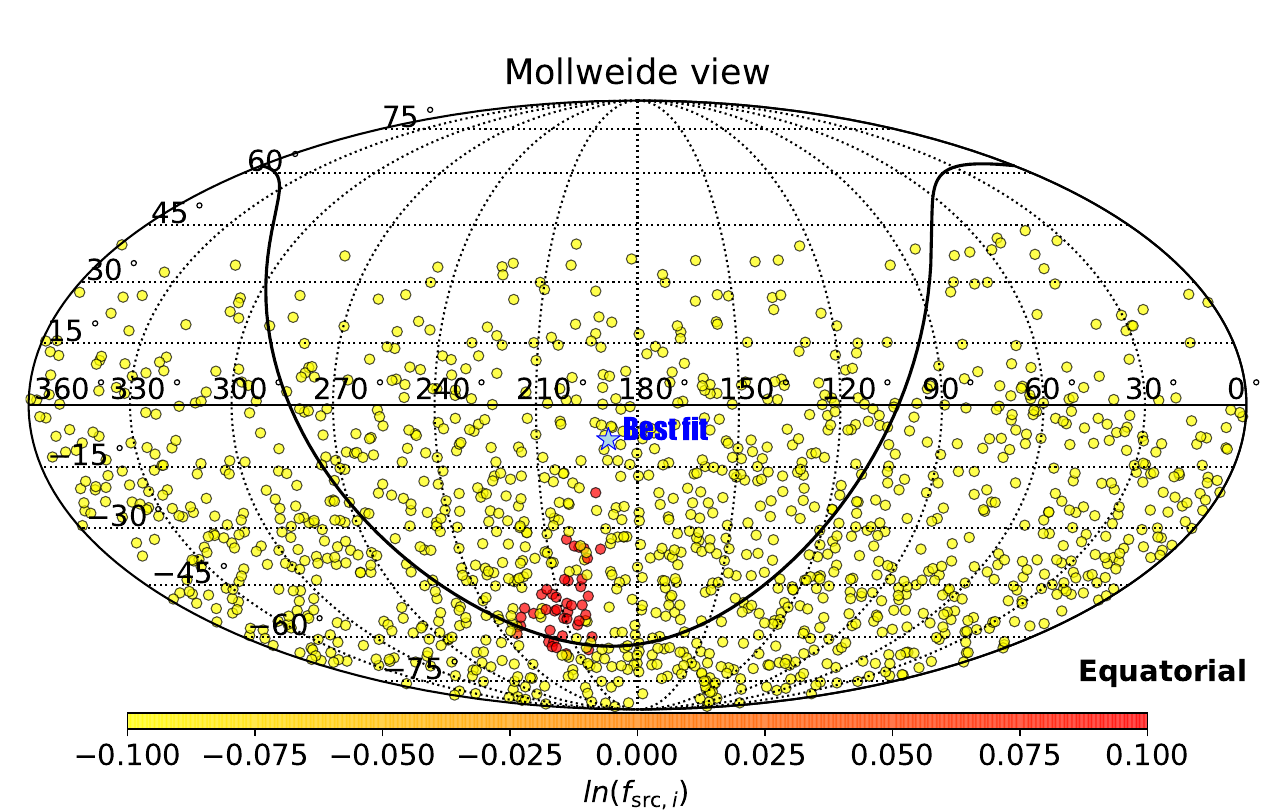} 
	\includegraphics[width=0.48\textwidth]{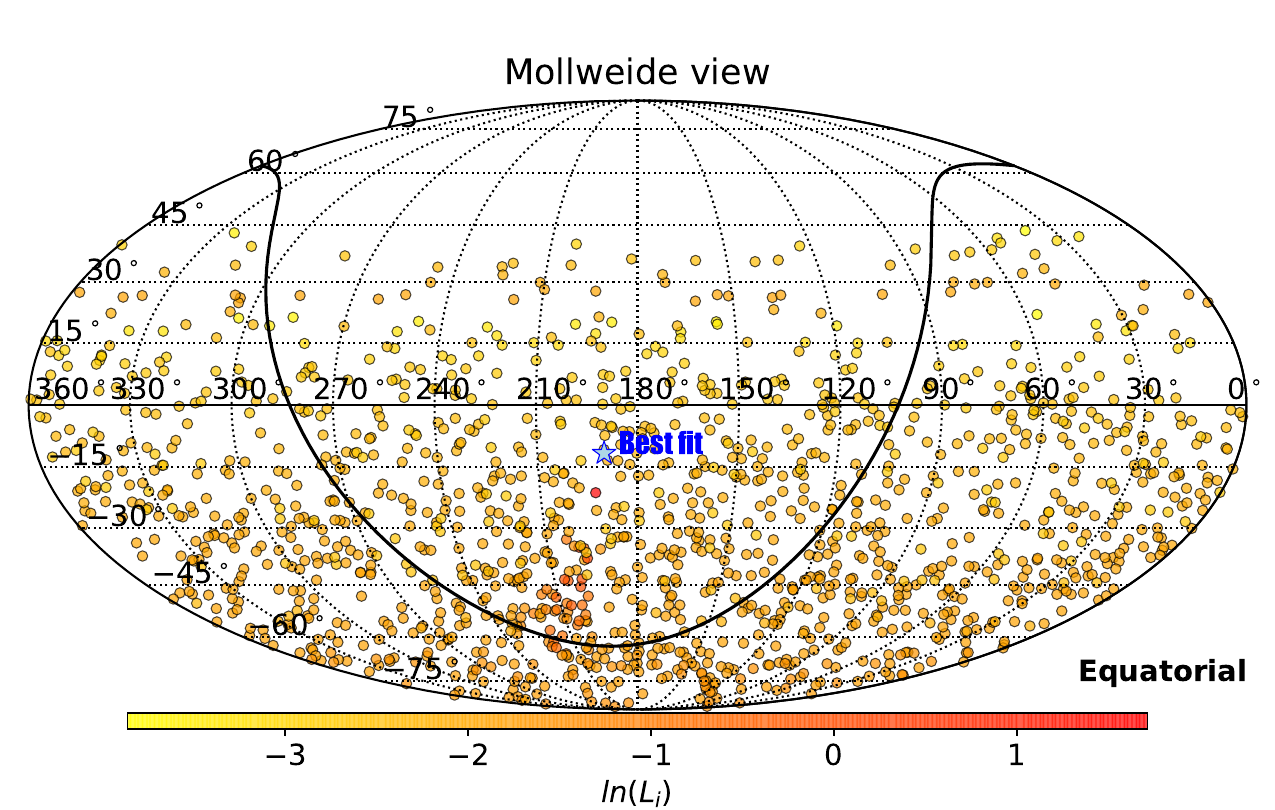} 
	\includegraphics[width=0.48\textwidth]{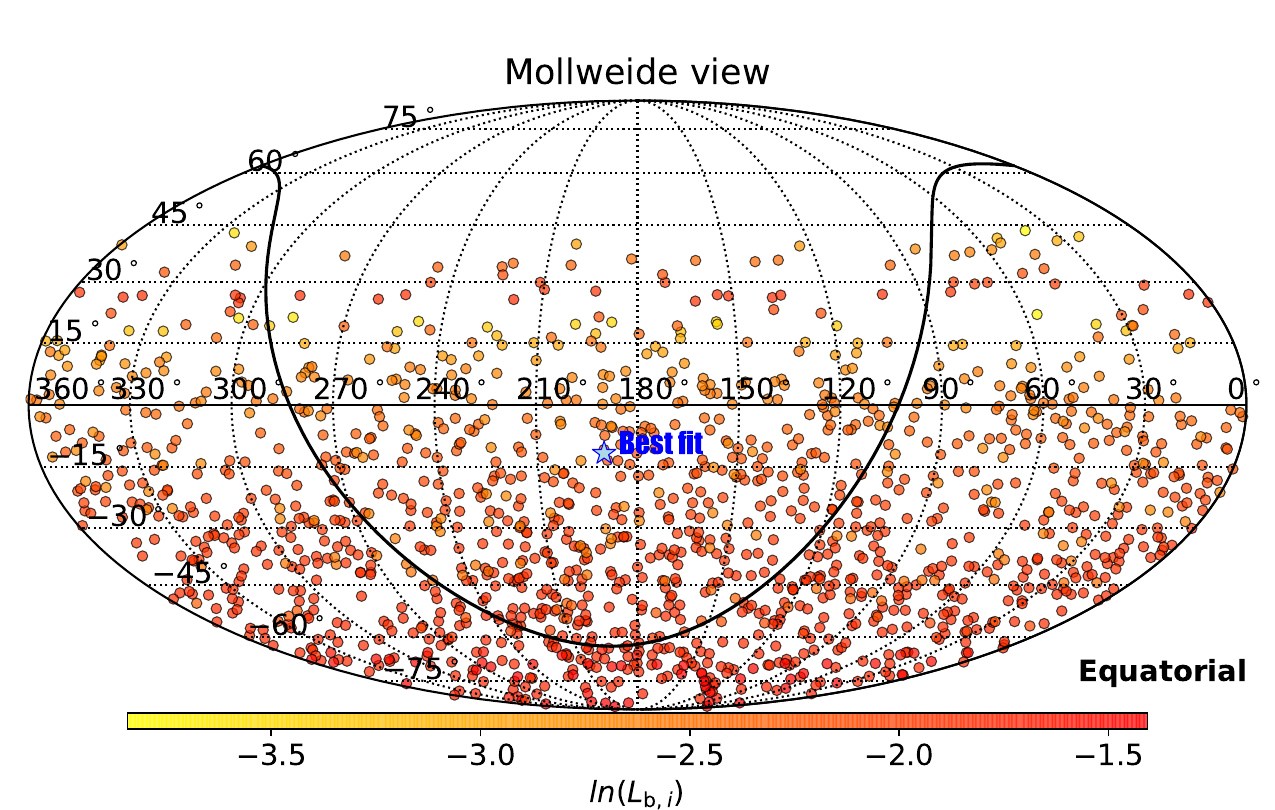} 
	\includegraphics[width=0.48\textwidth]{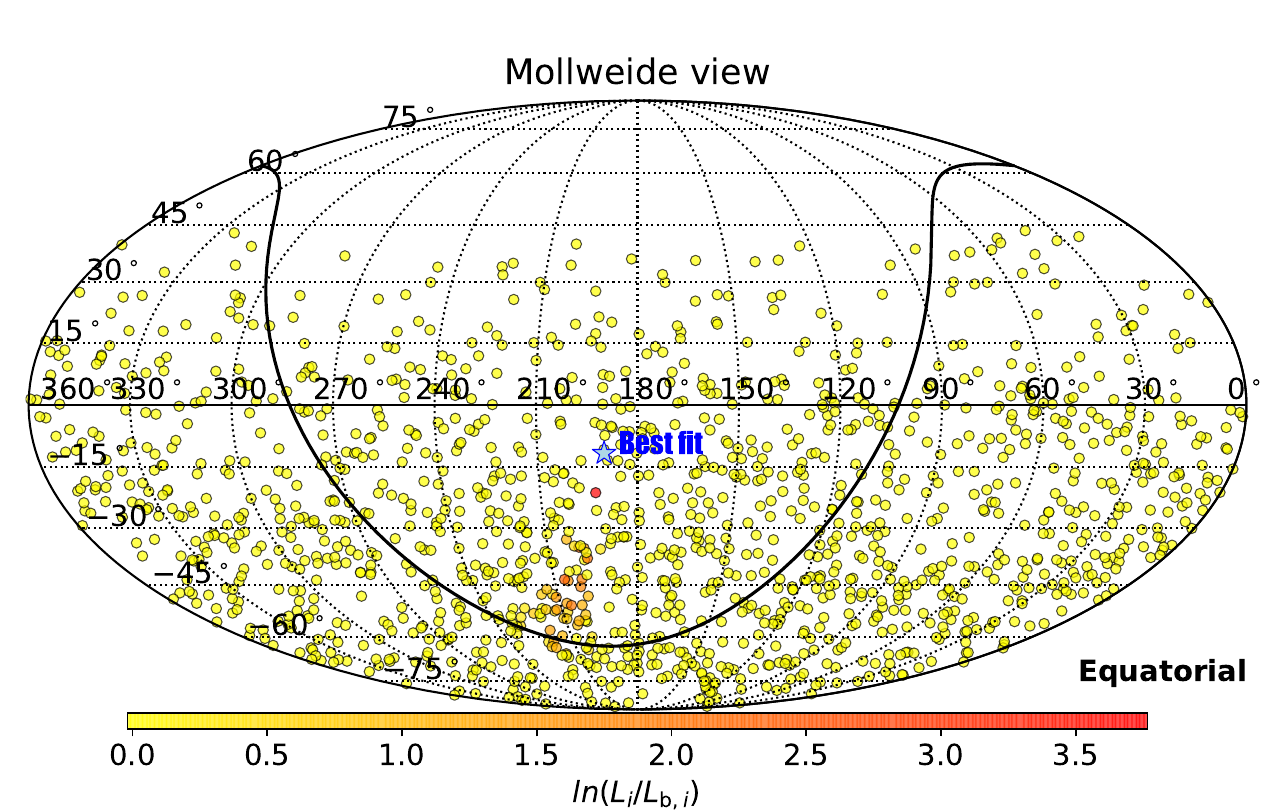} 

	\caption{\textbf{The value of $ln(f_{{\rm src},i})$, $ln(L_i)$, $ln(L_{{\rm b},i})$ and $ln(L_i/L_{{\rm b},i})$.}
		The value of $ln(f_{{\rm src},i})$, $ln(L_i)$, $ln(L_{{\rm b},i})$ and $ln(L_i/L_{{\rm b},i})$ for the PAO {\it Phase I} events with energies greater than 40 EeV, utilizing the 6 best-fit parameters. The blue star denotes the coordinates of the best-fitted source.}
	\label{fig:Li} 
\end{figure*}

Assuming that among the nearly isotropically distributed UHECRs there is only one UHECR multiplet originating from a point source, 
accounting for the background and varying exposure, a refined likelihood function for the i-th cosmic ray is written as 
\begin{eqnarray}
 &&L_i(E_i,\alpha_{i},\delta_{i}|\alpha_{\rm s},\delta_{\rm s}, A_{\rm reg}, A_{\rm ran}, \phi_{\rm reg},n_{\rm s}) \nonumber\\
 &=&\frac{n_{\rm s}}{ n_{\rm tot}} \frac{ f_{\rm src}(E_i,\alpha_{ i},\delta_{ i}|\alpha_{\rm s},\delta_{\rm s}, A_{\rm reg},A_{\rm ran}, \phi_{\rm reg}) \omega(\delta_i) }{\int_{4 \pi} f_{\rm src}(E_i,\alpha,\delta|\alpha_{\rm s},\delta_{\rm s}, A_{\rm reg},A_{\rm ran}, \phi_{\rm reg})\omega(\delta) d\Omega}+(1-\frac{n_{\rm s}}{n_{\rm tot}})\frac{ f_{\rm iso} \omega(\delta_i) }{\int_{4 \pi} f_{\rm iso} \omega (\delta) d\Omega}.
    \label{eq:lnLi}
\end{eqnarray}
where $n_{\rm s}$ represents the cosmic ray count in the UHECR multiplet, $n_{\rm tot}$ is the total count of UHECRs in the entire data set, $f_{\rm iso}=\frac{1}{4\pi}$ is the probability of an event originating from the isotropic background contribution.
Here, $\omega$ symbolizes the observatory's exposure varying as a function of the declination $\delta$ and the data type, which is computed for the vertical and inclined samples, respectively\cite{Auger:2022}. 
The likelihood function for the $i$-th cosmic ray being the background is written as 
\begin{equation}
    L_{{\rm b},i}=L_i(n_{\rm s}=0)=\frac{ f_{\rm iso} \omega(\delta_i) }{\int_{4 \pi} f_{\rm iso} \omega (\delta) d\Omega}
\end{equation}
The total likelihood function for the total UHECR data is written as
 \begin{equation}
   L(\alpha_{\rm s},\delta_{\rm s},A_{\rm reg},A_{\rm ran},\phi_{\rm reg}, n_{\rm s})= \prod^{n_{\rm tot}}_{i=0}  L_i(E_i, \alpha_{i},\beta_i|\alpha_{\rm s},\delta_{\rm s},A_{\rm reg},A_{\rm ran},\phi_{\rm reg}, n_{\rm s}).
    \label{eq:lnL}
\end{equation}
The value of $ln(f_{\rm src})$, $ln(L_{i})$,  $ln(L_{{\rm b},i})$ and $ln(L_i/L_{{\rm b},i})$ for each UHECR with energy larger than 40 EeV, utilizing the 6 best-fit parameters, are plotted in Figure \ref{fig:Li}. The plot of $ln(L_{{\rm b},i})$ shows that the vertical events and the inclined events follow the distribution of the exposures for the vertical and inclined events, respectively.
The plot of $ln(L_i/L_{{\rm b},i})$ shows that a group of UHECRs emerges from the nearly isotropic background.

The TS to compare the alternative hypothesis ($n_{\rm s}>0$) and the null hypothesis ($n_{\rm s}=0$) is defined as
\begin{equation}
    {\rm TS_{6para}}\equiv -2\times ln\left(\frac{L(n_{\rm s}=0)}{L(\alpha_{\rm s},\delta_{\rm s}, A_{\rm reg},A_{\rm ran},\phi_{\rm reg}, n_{\rm s})}\right).
\end{equation}

\subsubsection*{Calculating confidence level contours of the source coordinates}
The uncertainty in the reconstructed energies of the UHECRs impacts the positions of the apparent sources and consequently influences the measured $\theta_i$ and the distribution of $\theta_i$. 
To accommodate for uncertainties in the reconstructed energies, 
400 new datasets of UHECRs are generated, with the energies following Gaussian distributions around the reconstructed values, taking into account both statistical and systematic errors. The systematic error of the reconstructed energy does not influence the constrained coordinates of the source, but only on the constrained parameters of the magnetic field characteristics. 
By analyzing the new data sets with the 6-parameter maximum likelihood method, we derive the probability distribution for the 6 parameters $(\alpha_{\rm s},\delta_{\rm s},A_{\rm reg},A_{\rm ran},\phi_{\rm reg}, n_{\rm s})$ for each new data set.
Subsequently, $10^5$ samples of the six parameters are selected following the 6-parameter probability distributions for each new data set. 
The probability distribution of the source coordinates is derived by analyzing the $4\times10^7$ samples of the 6 parameters.
The contours of 68$\%$ and 95$\%$ confidence level  for the source coordinates are then obtained, as illustrated in Figure \ref{fig:4para}.

\subsubsection*{Simulating the background-only datasets}
To assess the significance of finding a UHECR multiplet or a multiplet associated with the Sombrero Galaxy, we simulate datasets of isotropically distributed UHECRs, ensuring that their distribution follows the PAO exposure, and no additional UHECR multiplet signals are included.
Firstly, we create a large sample of uniformly distributed declinations and selected 1,067 declinations for vertical events and 320 declinations for inclined events, using the PAO exposures as the selection weights, which accounts for the varying detection efficiency for the vertical and inclined events at different zenith angles. Next, we generate uniformly distributed R.A. for the vertical and inclined events,  respectively. For energies, we generate new energies following Gaussian distributions centered around the reconstructed energies with standard deviations that reflect both statistical and systematic errors. Then we randomly shuffled the new energies separately for vertical and inclined events and reassigned these shuffled energies to the corresponding simulated events, ensuring that the spatial distribution of events remained unchanged.

\subsubsection*{The likelihood ratio testing for the multiplet associated with the source}

When the source is fixed to be in the particular coordinates, the 6-parameter maximum likelihood method is transformed into a 4-parameter maximum likelihood method. The application of this 4-parameter approach allows the calculation of the optimal parameters and the corresponding errors. 
In order to assess the significance of a source contributing to a UHECR multiplet amidst the nearly isotropically distributed UHECR events, the likelihood ratio hypothesis testing is employed.
The TS to compare the alternative hypothesis ($n_{\rm s}>0$) and the null hypothesis ($n_{\rm s}=0$) at the given source coordinates ($\alpha_{\rm s},\delta_{\rm s}$)
is defined as
\begin{equation}
    {\rm TS_{source}}\equiv -2\times ln\left(\frac{L(n_{\rm s}=0)}{L(\alpha_{\rm s},\delta_{\rm s}| A_{\rm reg},A_{\rm ran},\phi_{\rm reg}, n_{\rm s})}\right).
\end{equation}

\subsubsection*{Source Catalogs}
We search for galaxy clusters, AGNs, starburst galaxies, and radio galaxies with jets/lobes within a conservative horizon distance of 36.6 Mpc\cite{Globus2023}, and then check whether there are any sources locate within the fitted 68$\%$ and 95$\%$ confidence level region of the source direction associated with the UHECR multiplet.
We searched for galaxy clusters from the X-ray-selected galaxy clusters catalog\cite{Koulouridis2021}. 
For AGNs, we adopted the same catalogs as referenced in the Pierre Auger Collaboration (2022)\cite{Auger:2022}.
This includes AGNs observed by {\it Swift}-BAT \cite{Oh2018} with the 14-195 keV flux exceeding $8.4\times10^{-12}{\rm erg~cm^{-2}~s^{-1}}$, gamma-ray selected AGNs from the {\it Fermi}-LAT 3FHL catalog \cite{Fermi2017} with an integral flux between 10 GeV and 1 TeV exceeding $3.3\times10^{-11}{\rm cm^{-2}~s^{-1}}$.
We searched for starburst galaxies with $L_{\rm FIR}>10^{10}L_\odot$ (with star formation rates above $1M_\odot \rm yr^{-1}$) from the starburst galaxy catalog used in the Pierre Auger Collaboration (2022)\cite{Auger:2022}, where $L_{\odot}=3.8\times10^{33}{\rm erg~s^{-1}}$ is the solar luminosity. 
For radio galaxies with jets/lobes, we first search for those classified as galaxies with jets/lobes in the bright master sample of radio galaxies as documented in van Velzen et al. (2012)\cite{vanVelzen2012}. 
As the bright master sample of radio galaxies excludes galaxies with a flux below 213 mJy at 1.4 GHz, it may overlook galaxies with jets or lobes exhibiting lower radio flux levels. 
Hence, we expand our search to identify galaxies with jets or lobes in the updated nearby galaxy catalog as of May 24, 2024\cite{Karachentsev2013}.
Based on the present observations, this investigation has identified 12 galaxy clusters, 31 starburst galaxies, 57 {\it Swift}-BAT AGNs, 3 {\it Fermi}-LAT AGNs, and 12 radio galaxies with jets/lobes located within the distance of 39.6 Mpc across the full sky. 
Except for the Sombrero Galaxy, none of these above sources is located within the 95$\%$ confidence region of the source direction.

Analysis of the mass composition of PAO events indicates a tendency towards heavier and less mixed compositions at the highest energies above $40$ EeV, where the fraction of CNO nuclei is greater than 50$\%$\cite{Mayotte2023}. It is possible that the observed nuclei have decayed from the heavier nuclei ejected from the source.
If the initial injected particles are iron nuclei, and the observed particles has $A_{\rm obs}\geq12$, the horizon distance is 36.6 Mpc.
Based on the present observations, there are 11 galaxy clusters, 30 starburst galaxies, 51 {\it Swift}-BAT AGNs, 3 {\it Fermi}-LAT AGNs, and 10 radio galaxies with jets/lobes located within the distance of 36.6 Mpc across the full sky. The conclusion does not change.

The horizon distance discussed above is a conservative estimation that ignores the deflection of the extragalactic magnetic field\cite{Globus2023}. The deflections by the extragalacitc magnetic field on UHECRs will enlarge the propagation length, thereby leading to an enhanced energy loss of the UHECRs and a consequent reduction in the horizon distance. 
If we assume the horizon distance of the 165 EeV event to be $13.5 ~{\rm Mpc}$ according to Bourriche $\&$ Capel (2023) \cite{Bourriche2023}, there are 2 galaxy clusters, 16 starburst galaxies, 13 {\it Swift}-BAT AGNs, 1 {\it Fermi}-LAT AGNs, and 2 radio galaxies with jets/lobes located within the distance of 13.5 Mpc across the sky. The Sombrero Galaxy still remains as the most probable source candidate.

\subsubsection*{The contribution from the 165 EeV UHECR}
The highest-energy UHECR in the UHECR multiplet with the energy of 165 EeV arrives closest to the direction of the Sombrero Galaxy comparing to the other UHECRs in the multiplet. In order to evaluate the influence of this particular event on the outcomes of our analysis, we have conducted a study in which the 165 EeV UHECR has been excluded from the data set. The fitted results for the UHECR multiplet and the multiplets associated with the Sombrero Galaxy and Centaurus A, based on the revised data set which excludes the 165 EeV event, are presented in Table \ref{tab:sources_r1}.

Upon removing the 165 EeV UHECR from the data set, the maximum TS value of the 6-parameter fitting decreases by $\sim$7, and the maximum TS value for the Sombrero Galaxy decreases by 3.6. 
This reduction indicates that the 165 EeV event contributes a small fraction to the overall significance of the association between the UHECR multiplet and the Sombrero Galaxy.
For Centaurus A, the maximum TS value exhibits little change after the exclusion of the 165 EeV event. This stability suggests that the 165 EeV event does not influence the probability of the association between the UHECR multiplet and Centaurus A.

\newpage


\begin{table*} 
	\centering
	\caption{\textbf{Summary of fitted parameters, the maximum TS value, the global P value and significance.}
		We report the best-fit parameters with 1-$\sigma$ errors for the multiplet found in the 6-parameter search, and the multiplet associated with the Sombrero Galaxy and Centaurus A, the corresponding maximum TS values, the degree of freedom, the global P value and the global significance, using the PAO {\it Phase I} data with energies greater than 40 EeV, adopting the reconstructed energy without taking into account the energy errors.
        }
	\label{tab:sources} 
	
	\begin{tabular}{lcccc} 
		\\
		\hline
			&Best fit &The Sombrero Galaxy& Centaurus A  \\   
		 \hline
		$\alpha_{\rm s}$&$188.7^{\circ+1.8^\circ}_{~-1.8^\circ}$ &$190.0^\circ$&$201.4^\circ$\\
		 $\delta_{\rm s}$&$-8.3^{\circ+1.8^\circ}_{~-1.9^\circ}$ &$-11.6^\circ$&$-43.0^\circ$\\
		$A_{\rm reg}$& $22.8^{\circ+1.3^\circ}_{~-1.3^\circ}$&$20.7^{\circ+0.57^\circ}_{~-0.58^\circ}$&$2.8^{\circ+5.6^\circ}_{~-2.8^\circ}$\\
		 $A_{\rm ran}$&$2.5^{\circ+0.47^\circ}_{~-0.38^\circ}$&$2.7^{\circ+0.56^\circ}_{~-0.40^\circ}$&$16.9^{\circ+6.8^\circ}_{~-6.6^\circ}$\\
		$\phi_{\rm reg}$ & $197.5^{\circ+2.7^\circ}_{~-2.6^\circ}$& $197.7^{\circ+1.7^\circ}_{~-1.6^\circ}$& $57.3^{\circ+302.7^\circ}_{~-57.3^\circ}$\\
		$n_{\rm s}$&$25.2^{+6.1}_{-6.9}$&$25.7^{+6.2}_{-7.0}$&$91.7^{+50.7}_{-43.1}$\\
		$\rm TS$&41.1&37.6&18.2\\
        degree of freedom& 6&4&4\\
        global P value&$2.4\times10^{-4}$&$9.8\times10^{-4}$&0.25\\
        global significance&3.7~$\sigma$&$3.3~\sigma$&$1.1~\sigma$\\
		\hline
	\end{tabular}
\end{table*}

\begin{table*} 
	\centering
	\caption{\textbf{Fitted parameters and the maximum TS value via removing the 165 EeV event.}
	 We report the fitted parameters and the maximum TS value for the 6-parameter best-fit result and for the Sombrero Galaxy and Centaurus A adopting the PAO {\it Phase I} data with energy larger than 40 EeV without taking into account the energy errors, and excluding the 165 EeV event.}
	\label{tab:sources_r1} 

	\begin{tabular}{lcccr} 
		\\
		\hline
			&Best Fit &The Sombrero Galaxy& Centaurus A  \\   
		 \hline
		$\alpha_{\rm s}$ &$190.2^{\circ+3.8^\circ}_{~-4.4^\circ}$&$190.0^\circ$&$201.4^\circ$\\
		 $\delta_{\rm s}$&$-10.9^{\circ+4.4^\circ}_{~-4.6^\circ}$ &$-11.6^\circ$&$-43.0^\circ$\\
		$A_{\rm reg}$ &$21.3^{\circ+2.6^\circ}_{~-2.6^\circ}$&$20.9^{\circ+0.6^\circ}_{~-0.6^\circ}$&$2.0^{\circ+5.4^\circ}_{~-2.0^\circ}$\\
		 $A_{\rm ran}$&$2.5^{\circ+0.5^\circ}_{~-0.4^\circ}$&$2.5^{\circ+0.5^\circ}_{~-0.4^\circ}$&$15.9^{\circ+6.2^\circ}_{~-5.6^\circ}$\\
		$\phi_{\rm reg}$&$197.1^{\circ+4.6^\circ}_{~-4.4^\circ}$ & $197.6^{\circ+1.7^\circ}_{~-1.6^\circ}$&$56.4^{\circ+303.6^\circ}_{~-56.4^\circ}$\\
		$n_{\rm s}$&$23.9^{+6.0}_{-6.9}$&$24.0^{+6.0}_{-6.8}$&$85.7^{+37.4}_{-45.7}$\\
		$\rm TS$&34.02&34.00&18.24\\
        degree of freedom &6&4&4\\
		\hline
	\end{tabular}
\end{table*}

\newpage

\clearpage

\bibliography{UHECR.bib}

\providecommand{\noopsort}[1]{}\providecommand{\singleletter}[1]{#1}%
\begin{thebibliography}{10}
\expandafter\ifx\csname url\endcsname\relax
  \def\url#1{\texttt{#1}}\fi
\expandafter\ifx\csname urlprefix\endcsname\relax\def\urlprefix{URL }\fi
\providecommand{\bibinfo}[2]{#2}
\providecommand{\eprint}[2][]{\url{#2}}

\bibitem{Sigl2001}
\bibinfo{author}{{Sigl}, G.}
\newblock \bibinfo{title}{{Ultrahigh-Energy Cosmic Rays: Physics and
  Astrophysics at Extreme Energies}}.
\newblock \emph{\bibinfo{journal}{Science}} \textbf{\bibinfo{volume}{291}},
  \bibinfo{pages}{73--79} (\bibinfo{year}{2001}).
\newblock \eprint{astro-ph/0104291}.

\bibitem{Auger:2007Sci}
\bibinfo{author}{{Pierre Auger Collaboration}} \emph{et~al.}
\newblock \bibinfo{title}{{Correlation of the Highest-Energy Cosmic Rays with
  Nearby Extragalactic Objects}}.
\newblock \emph{\bibinfo{journal}{Science}} \textbf{\bibinfo{volume}{318}},
  \bibinfo{pages}{938} (\bibinfo{year}{2007}).
\newblock \eprint{0711.2256}.

\bibitem{TA:2014}
\bibinfo{author}{{Abbasi}, R.~U.} \emph{et~al.}
\newblock \bibinfo{title}{{Indications of Intermediate-scale Anisotropy of
  Cosmic Rays with Energy Greater Than 57 EeV in the Northern Sky Measured with
  the Surface Detector of the Telescope Array Experiment}}.
\newblock \emph{\bibinfo{journal}{\apjl}} \textbf{\bibinfo{volume}{790}},
  \bibinfo{pages}{L21} (\bibinfo{year}{2014}).
\newblock \eprint{1404.5890}.

\bibitem{Auger:2017Sci}
\bibinfo{author}{{Pierre Auger Collaboration}} \emph{et~al.}
\newblock \bibinfo{title}{{Observation of a large-scale anisotropy in the
  arrival directions of cosmic rays above 8 {\texttimes} {}10$^{18}$ eV}}.
\newblock \emph{\bibinfo{journal}{Science}} \textbf{\bibinfo{volume}{357}},
  \bibinfo{pages}{1266--1270} (\bibinfo{year}{2017}).
\newblock \eprint{1709.07321}.

\bibitem{Auger:2020prl}
\bibinfo{author}{{Aab}, A.} \emph{et~al.}
\newblock \bibinfo{title}{{Features of the Energy Spectrum of Cosmic Rays above
  2.5 {\texttimes}{}10$^{18}$ eV Using the Pierre Auger Observatory}}.
\newblock \emph{\bibinfo{journal}{\prl}} \textbf{\bibinfo{volume}{125}},
  \bibinfo{pages}{121106} (\bibinfo{year}{2020}).
\newblock \eprint{2008.06488}.

\bibitem{Auger:2022}
\bibinfo{author}{{Abreu}, P.} \emph{et~al.}
\newblock \bibinfo{title}{{Arrival Directions of Cosmic Rays above 32 EeV from
  Phase One of the Pierre Auger Observatory}}.
\newblock \emph{\bibinfo{journal}{\apj}} \textbf{\bibinfo{volume}{935}},
  \bibinfo{pages}{170} (\bibinfo{year}{2022}).
\newblock \eprint{2206.13492}.

\bibitem{TA:2024}
\bibinfo{author}{{Telescope Array Collaboration}} \emph{et~al.}
\newblock \bibinfo{title}{{An extremely energetic cosmic ray observed by a
  surface detector array}}.
\newblock \emph{\bibinfo{journal}{Science}} \textbf{\bibinfo{volume}{382}},
  \bibinfo{pages}{903--907} (\bibinfo{year}{2023}).
\newblock \eprint{2311.14231}.

\bibitem{Golup:2009}
\bibinfo{author}{{G. Golup, D. Harari, S. Mollerach, and E. Roulet}}.
\newblock \bibinfo{title}{{Source position reconstruction and constraints on
  the galactic magnetic field from ultra-high energy cosmic rays}}.
\newblock \emph{\bibinfo{journal}{Astropart. Phys.}}
  \textbf{\bibinfo{volume}{32}}, \bibinfo{pages}{269--277}
  (\bibinfo{year}{2009}).

\bibitem{Harari:2001}
\bibinfo{author}{{D. Harari, S. Mollerach, E. Roulet, and F. S\'anchez}}.
\newblock \bibinfo{title}{{Lensing of ultra-high energy cosmic rays in
  turbulent magnetic fields}}.
\newblock \emph{\bibinfo{journal}{Journal of High Energy Physics}}
  \textbf{\bibinfo{volume}{3}}, \bibinfo{pages}{45} (\bibinfo{year}{2002}).

\bibitem{Tinyakov:2005}
\bibinfo{author}{{P. G. Tinyakov and I. I. Tkachev}}.
\newblock \bibinfo{title}{{Deflections of cosmic rays in a random component of
  the Galactic magnetic field}}.
\newblock \emph{\bibinfo{journal}{Astroparticle Physics}}
  \textbf{\bibinfo{volume}{24}}, \bibinfo{pages}{32} (\bibinfo{year}{2005}).

\bibitem{Auger:2012multiplets}
\bibinfo{author}{{Pierre Auger Collaboration}} \emph{et~al.}
\newblock \bibinfo{title}{{Search for signatures of magnetically-induced
  alignment in the arrival directions measured by the Pierre Auger
  Observatory}}.
\newblock \emph{\bibinfo{journal}{Astroparticle Physics}}
  \textbf{\bibinfo{volume}{35}}, \bibinfo{pages}{354--361}
  (\bibinfo{year}{2012}).
\newblock \eprint{1111.2472}.

\bibitem{He:2016}
\bibinfo{author}{{Hao-Ning He, Alexander Kusenko, Shigehiro Nagataki, Bin-Bin
  Zhang, Rui-Zhi Yang, and Yi-Zhong Fan}}.
\newblock \bibinfo{title}{{Monte Carlo Bayesian search for the plausible source
  of the Telescope Array hotspot}}.
\newblock \emph{\bibinfo{journal}{Phys. Rev. D}} \textbf{\bibinfo{volume}{93}},
  \bibinfo{pages}{043011} (\bibinfo{year}{2016}).

\bibitem{Auger:2015multiplets}
\bibinfo{author}{{Aab}, A.} \emph{et~al.}
\newblock \bibinfo{title}{{Search for patterns by combining cosmic-ray energy
  and arrival directions at the Pierre Auger Observatory}}.
\newblock \emph{\bibinfo{journal}{European Physical Journal C}}
  \textbf{\bibinfo{volume}{75}}, \bibinfo{pages}{269} (\bibinfo{year}{2015}).
\newblock \eprint{1410.0515}.

\bibitem{Auger:2020}
\bibinfo{author}{{A. Aab et al. (The Pierre Auger Collaboration)}}.
\newblock \bibinfo{title}{{Search for magnetically-induced signatures in the
  arrival directions of ultra-high-energy cosmic rays measured at the Pierre
  Auger Observatory}}.
\newblock \emph{\bibinfo{journal}{Journal of Cosmology and Astroparticle
  Physics}} \textbf{\bibinfo{volume}{06}}, \bibinfo{pages}{017 (2020)}
  (\bibinfo{year}{2020}).

\bibitem{TA:2020}
\bibinfo{author}{{Abbasi}, R.~U.} \emph{et~al.}
\newblock \bibinfo{title}{{Evidence for a Supergalactic Structure of Magnetic
  Deflection Multiplets of Ultra-high-energy Cosmic Rays}}.
\newblock \emph{\bibinfo{journal}{\apj}} \textbf{\bibinfo{volume}{899}},
  \bibinfo{pages}{86} (\bibinfo{year}{2020}).

\bibitem{Mayotte2023}
\bibinfo{author}{Mayotte, E.~W.} \emph{et~al.}
\newblock \bibinfo{title}{{Measurement of the mass composition of
  ultra-high-energy cosmic rays at the Pierre Auger Observatory}}.
\newblock \emph{\bibinfo{journal}{PoS}} \textbf{\bibinfo{volume}{ICRC2023}},
  \bibinfo{pages}{365} (\bibinfo{year}{2023}).

\bibitem{Auger:2023catalog}
\bibinfo{author}{{Abdul Halim}, A.} \emph{et~al.}
\newblock \bibinfo{title}{{A Catalog of the Highest-energy Cosmic Rays Recorded
  during Phase I of Operation of the Pierre Auger Observatory}}.
\newblock \emph{\bibinfo{journal}{\apjs}} \textbf{\bibinfo{volume}{264}},
  \bibinfo{pages}{50} (\bibinfo{year}{2023}).

\bibitem{G1966}
\bibinfo{author}{{Greisen}, K.}
\newblock \bibinfo{title}{{End to the Cosmic-Ray Spectrum?}}
\newblock \emph{\bibinfo{journal}{\prl}} \textbf{\bibinfo{volume}{16}},
  \bibinfo{pages}{748--750} (\bibinfo{year}{1966}).

\bibitem{ZK1966}
\bibinfo{author}{{Zatsepin}, G.~T.} \& \bibinfo{author}{{Kuz'min}, V.~A.}
\newblock \bibinfo{title}{{Upper Limit of the Spectrum of Cosmic Rays}}.
\newblock \emph{\bibinfo{journal}{Soviet Journal of Experimental and
  Theoretical Physics Letters}} \textbf{\bibinfo{volume}{4}},
  \bibinfo{pages}{78} (\bibinfo{year}{1966}).

\bibitem{Globus2023}
\bibinfo{author}{{Globus}, N.}, \bibinfo{author}{{Fedynitch}, A.} \&
  \bibinfo{author}{{Blandford}, R.~D.}
\newblock \bibinfo{title}{{Treasure Maps for Detections of Extreme Energy
  Cosmic Rays}}.
\newblock \emph{\bibinfo{journal}{\apj}} \textbf{\bibinfo{volume}{945}},
  \bibinfo{pages}{12} (\bibinfo{year}{2023}).
\newblock \eprint{2210.15885}.

\bibitem{Anchordoqui:2018qom}
\bibinfo{author}{Anchordoqui, L.~A.}
\newblock \bibinfo{title}{{Ultra-High-Energy Cosmic Rays}}.
\newblock \emph{\bibinfo{journal}{Phys. Rept.}} \textbf{\bibinfo{volume}{801}},
  \bibinfo{pages}{1--93} (\bibinfo{year}{2019}).
\newblock \eprint{1807.09645}.

\bibitem{Koulouridis2021}
\bibinfo{author}{{Koulouridis}, E.} \emph{et~al.}
\newblock \bibinfo{title}{{The X-CLASS survey: A catalogue of 1646
  X-ray-selected galaxy clusters up to z {\ensuremath{\sim}} 1.5}}.
\newblock \emph{\bibinfo{journal}{\aap}} \textbf{\bibinfo{volume}{652}},
  \bibinfo{pages}{A12} (\bibinfo{year}{2021}).
\newblock \eprint{2104.06617}.

\bibitem{Baumgartner2013}
\bibinfo{author}{{Baumgartner}, W.~H.} \emph{et~al.}
\newblock \bibinfo{title}{{The 70 Month Swift-BAT All-sky Hard X-Ray Survey}}.
\newblock \emph{\bibinfo{journal}{\apjs}} \textbf{\bibinfo{volume}{207}},
  \bibinfo{pages}{19} (\bibinfo{year}{2013}).
\newblock \eprint{1212.3336}.

\bibitem{Ajello2017}
\bibinfo{author}{{Ajello}, M.} \emph{et~al.}
\newblock \bibinfo{title}{{3FHL: The Third Catalog of Hard Fermi-LAT Sources}}.
\newblock \emph{\bibinfo{journal}{\apjs}} \textbf{\bibinfo{volume}{232}},
  \bibinfo{pages}{18} (\bibinfo{year}{2017}).
\newblock \eprint{1702.00664}.

\bibitem{vanVelzen2012}
\bibinfo{author}{{van Velzen}, S.}, \bibinfo{author}{{Falcke}, H.},
  \bibinfo{author}{{Schellart}, P.}, \bibinfo{author}{{Nierstenh{\"o}fer}, N.}
  \& \bibinfo{author}{{Kampert}, K.-H.}
\newblock \bibinfo{title}{{Radio galaxies of the local universe. All-sky
  catalog, luminosity functions, and clustering}}.
\newblock \emph{\bibinfo{journal}{\aap}} \textbf{\bibinfo{volume}{544}},
  \bibinfo{pages}{A18} (\bibinfo{year}{2012}).
\newblock \eprint{1206.0031}.

\bibitem{Karachentsev2013}
\bibinfo{author}{{Karachentsev}, I.~D.}, \bibinfo{author}{{Makarov}, D.~I.} \&
  \bibinfo{author}{{Kaisina}, E.~I.}
\newblock \bibinfo{title}{{Updated Nearby Galaxy Catalog}}.
\newblock \emph{\bibinfo{journal}{\aj}} \textbf{\bibinfo{volume}{145}},
  \bibinfo{pages}{101} (\bibinfo{year}{2013}).
\newblock \eprint{1303.5328}.

\bibitem{Jansson2012a}
\bibinfo{author}{{Jansson}, R.} \& \bibinfo{author}{{Farrar}, G.~R.}
\newblock \bibinfo{title}{{A New Model of the Galactic Magnetic Field}}.
\newblock \emph{\bibinfo{journal}{\apj}} \textbf{\bibinfo{volume}{757}},
  \bibinfo{pages}{14} (\bibinfo{year}{2012}).
\newblock \eprint{1204.3662}.

\bibitem{Jansson2012b}
\bibinfo{author}{{Jansson}, R.} \& \bibinfo{author}{{Farrar}, G.~R.}
\newblock \bibinfo{title}{{The Galactic Magnetic Field}}.
\newblock \emph{\bibinfo{journal}{\apjl}} \textbf{\bibinfo{volume}{761}},
  \bibinfo{pages}{L11} (\bibinfo{year}{2012}).
\newblock \eprint{1210.7820}.

\bibitem{Batista2022}
\bibinfo{author}{{Alves Batista}, R.} \emph{et~al.}
\newblock \bibinfo{title}{{CRPropa 3.2 - an advanced framework for high-energy
  particle propagation in extragalactic and galactic spaces}}.
\newblock \emph{\bibinfo{journal}{\jcap}} \textbf{\bibinfo{volume}{2022}},
  \bibinfo{pages}{035} (\bibinfo{year}{2022}).
\newblock \eprint{2208.00107}.

\bibitem{McQuinn2016}
\bibinfo{author}{{McQuinn}, K. B.~W.}, \bibinfo{author}{{Skillman}, E.~D.},
  \bibinfo{author}{{Dolphin}, A.~E.}, \bibinfo{author}{{Berg}, D.} \&
  \bibinfo{author}{{Kennicutt}, R.}
\newblock \bibinfo{title}{{The Distance to M104}}.
\newblock \emph{\bibinfo{journal}{\aj}} \textbf{\bibinfo{volume}{152}},
  \bibinfo{pages}{144} (\bibinfo{year}{2016}).

\bibitem{Kormendy1988}
\bibinfo{author}{{Kormendy}, J.}
\newblock \bibinfo{title}{{Evidence for a Central Dark Mass in NGC 4594 (The
  Sombrero Galaxy)}}.
\newblock \emph{\bibinfo{journal}{\apj}} \textbf{\bibinfo{volume}{335}},
  \bibinfo{pages}{40} (\bibinfo{year}{1988}).

\bibitem{Kormendy1996}
\bibinfo{author}{{Kormendy}, J.} \emph{et~al.}
\newblock \bibinfo{title}{{Hubble Space Telescope Spectroscopic Evidence for a
  1 X 10 9 M$_{sun}$ Black Hole in NGC 4594}}.
\newblock \emph{\bibinfo{journal}{\apjl}} \textbf{\bibinfo{volume}{473}},
  \bibinfo{pages}{L91} (\bibinfo{year}{1996}).

\bibitem{Neronov2009}
\bibinfo{author}{{Neronov}, A.~Y.}, \bibinfo{author}{{Semikoz}, D.~V.} \&
  \bibinfo{author}{{Tkachev}, I.~I.}
\newblock \bibinfo{title}{{Ultra-high energy cosmic ray production in the polar
  cap regions of black hole magnetospheres}}.
\newblock \emph{\bibinfo{journal}{New Journal of Physics}}
  \textbf{\bibinfo{volume}{11}}, \bibinfo{pages}{065015}
  (\bibinfo{year}{2009}).
\newblock \eprint{0712.1737}.

\bibitem{Tursunov2020}
\bibinfo{author}{{Tursunov}, A.}, \bibinfo{author}{{Stuchl{\'\i}k}, Z.},
  \bibinfo{author}{{Kolo{\v{s}}}, M.}, \bibinfo{author}{{Dadhich}, N.} \&
  \bibinfo{author}{{Ahmedov}, B.}
\newblock \bibinfo{title}{{Supermassive Black Holes as Possible Sources of
  Ultrahigh-energy Cosmic Rays}}.
\newblock \emph{\bibinfo{journal}{\apj}} \textbf{\bibinfo{volume}{895}},
  \bibinfo{pages}{14} (\bibinfo{year}{2020}).
\newblock \eprint{2004.07907}.

\bibitem{Gallimore2006}
\bibinfo{author}{{Gallimore}, J.~F.}, \bibinfo{author}{{Axon}, D.~J.},
  \bibinfo{author}{{O'Dea}, C.~P.}, \bibinfo{author}{{Baum}, S.~A.} \&
  \bibinfo{author}{{Pedlar}, A.}
\newblock \bibinfo{title}{{A Survey of Kiloparsec-Scale Radio Outflows in
  Radio-Quiet Active Galactic Nuclei}}.
\newblock \emph{\bibinfo{journal}{\aj}} \textbf{\bibinfo{volume}{132}},
  \bibinfo{pages}{546--569} (\bibinfo{year}{2006}).
\newblock \eprint{astro-ph/0604219}.

\bibitem{Li2011}
\bibinfo{author}{{Li}, Z.} \emph{et~al.}
\newblock \bibinfo{title}{{X-ray Emission from the Sombrero Galaxy: A
  Galactic-scale Outflow}}.
\newblock \emph{\bibinfo{journal}{\apj}} \textbf{\bibinfo{volume}{730}},
  \bibinfo{pages}{84} (\bibinfo{year}{2011}).
\newblock \eprint{1009.5767}.

\bibitem{Hada2013}
\bibinfo{author}{{Hada}, K.} \emph{et~al.}
\newblock \bibinfo{title}{{Evidence for a Nuclear Radio Jet and its Structure
  down to lsim100 Schwarzschild Radii in the Center of the Sombrero Galaxy (M
  104, NGC 4594)}}.
\newblock \emph{\bibinfo{journal}{\apj}} \textbf{\bibinfo{volume}{779}},
  \bibinfo{pages}{6} (\bibinfo{year}{2013}).
\newblock \eprint{1310.0488}.

\bibitem{Mezcua2014}
\bibinfo{author}{{Mezcua}, M.} \& \bibinfo{author}{{Prieto}, M.~A.}
\newblock \bibinfo{title}{{Evidence of Parsec-scale Jets in Low-luminosity
  Active Galactic Nuclei}}.
\newblock \emph{\bibinfo{journal}{\apj}} \textbf{\bibinfo{volume}{787}},
  \bibinfo{pages}{62} (\bibinfo{year}{2014}).
\newblock \eprint{1403.6675}.

\bibitem{Yang2024}
\bibinfo{author}{{Yang}, Y.} \emph{et~al.}
\newblock \bibinfo{title}{{CHANG-ES. XXX. 10 kpc Radio Lobes in the Sombrero
  Galaxy}}.
\newblock \emph{\bibinfo{journal}{\apj}} \textbf{\bibinfo{volume}{966}},
  \bibinfo{pages}{213} (\bibinfo{year}{2024}).
\newblock \eprint{2403.16682}.

\bibitem{Yan2024}
\bibinfo{author}{{Yan}, X.} \emph{et~al.}
\newblock \bibinfo{title}{{Multifrequency Very Long Baseline Interferometry
  Imaging of the Subparsec-scale Jet in the Sombrero Galaxy (M104)}}.
\newblock \emph{\bibinfo{journal}{\apj}} \textbf{\bibinfo{volume}{965}},
  \bibinfo{pages}{128} (\bibinfo{year}{2024}).
\newblock \eprint{2403.04215}.

\bibitem{Alves2018}
\bibinfo{author}{{Alves}, E.~P.}, \bibinfo{author}{{Zrake}, J.} \&
  \bibinfo{author}{{Fiuza}, F.}
\newblock \bibinfo{title}{{Efficient Nonthermal Particle Acceleration by the
  Kink Instability in Relativistic Jets}}.
\newblock \emph{\bibinfo{journal}{\prl}} \textbf{\bibinfo{volume}{121}},
  \bibinfo{pages}{245101} (\bibinfo{year}{2018}).
\newblock \eprint{1810.05154}.

\bibitem{Mattews2019}
\bibinfo{author}{{Matthews}, J.~H.}, \bibinfo{author}{{Bell}, A.~R.},
  \bibinfo{author}{{Blundell}, K.~M.} \& \bibinfo{author}{{Araudo}, A.~T.}
\newblock \bibinfo{title}{{Ultrahigh energy cosmic rays from shocks in the
  lobes of powerful radio galaxies}}.
\newblock \emph{\bibinfo{journal}{\mnras}} \textbf{\bibinfo{volume}{482}},
  \bibinfo{pages}{4303--4321} (\bibinfo{year}{2019}).
\newblock \eprint{1810.12350}.

\bibitem{Hillas1984}
\bibinfo{author}{{Hillas}, A.~M.}
\newblock \bibinfo{title}{{The Origin of Ultra-High-Energy Cosmic Rays}}.
\newblock \emph{\bibinfo{journal}{\araa}} \textbf{\bibinfo{volume}{22}},
  \bibinfo{pages}{425--444} (\bibinfo{year}{1984}).

\bibitem{Wykes2013}
\bibinfo{author}{{Wykes}, S.} \emph{et~al.}
\newblock \bibinfo{title}{{Mass entrainment and turbulence-driven acceleration
  of ultra-high energy cosmic rays in Centaurus A}}.
\newblock \emph{\bibinfo{journal}{\aap}} \textbf{\bibinfo{volume}{558}},
  \bibinfo{pages}{A19} (\bibinfo{year}{2013}).
\newblock \eprint{1305.2761}.

\bibitem{Han2017}
\bibinfo{author}{{Han}, J.~L.}
\newblock \bibinfo{title}{{Observing Interstellar and Intergalactic Magnetic
  Fields}}.
\newblock \emph{\bibinfo{journal}{\araa}} \textbf{\bibinfo{volume}{55}},
  \bibinfo{pages}{111--157} (\bibinfo{year}{2017}).

\bibitem{Xu2024}
\bibinfo{author}{{Xu}, J.} \& \bibinfo{author}{{Han}, J.~L.}
\newblock \bibinfo{title}{{The Huge Magnetic Toroids in the Milky Way Halo}}.
\newblock \emph{\bibinfo{journal}{\apj}} \textbf{\bibinfo{volume}{966}},
  \bibinfo{pages}{240} (\bibinfo{year}{2024}).
\newblock \eprint{2404.02038}.

\bibitem{GRAND2020}
\bibinfo{author}{{{\'A}lvarez-Mu{\~n}iz}, J.} \emph{et~al.}
\newblock \bibinfo{title}{{The Giant Radio Array for Neutrino Detection
  (GRAND): Science and design}}.
\newblock \emph{\bibinfo{journal}{Science China Physics, Mechanics, and
  Astronomy}} \textbf{\bibinfo{volume}{63}}, \bibinfo{pages}{219501}
  (\bibinfo{year}{2020}).
\newblock \eprint{1810.09994}.

\bibitem{AugerPrime2022}
\bibinfo{author}{{Stasielak}, J.}
\newblock \bibinfo{title}{{AugerPrime - The upgrade of the Pierre Auger
  Observatory}}.
\newblock \emph{\bibinfo{journal}{International Journal of Modern Physics A}}
  \textbf{\bibinfo{volume}{37}}, \bibinfo{pages}{2240012--251}
  (\bibinfo{year}{2022}).
\newblock \eprint{2110.09487}.

\bibitem{Xia:2022uua}
\bibinfo{author}{Xia, Z.-Q.}, \bibinfo{author}{Wang, Y.},
  \bibinfo{author}{Yuan, Q.} \& \bibinfo{author}{Fan, Y.-Z.}
\newblock \bibinfo{title}{{A delayed 400 GeV photon from GRB 221009A and
  implication on the intergalactic magnetic field}}.
\newblock \emph{\bibinfo{journal}{Nature Commun.}}
  \textbf{\bibinfo{volume}{15}}, \bibinfo{pages}{4280} (\bibinfo{year}{2024}).
\newblock \eprint{2210.13052}.

\bibitem{Fisher1953}
\bibinfo{author}{{Fisher}, R.}
\newblock \bibinfo{title}{{Dispersion on a Sphere}}.
\newblock \emph{\bibinfo{journal}{Proceedings of the Royal Society of London
  Series A}} \textbf{\bibinfo{volume}{217}}, \bibinfo{pages}{295--305}
  (\bibinfo{year}{1953}).

\bibitem{Oh2018}
\bibinfo{author}{{Oh}, K.} \emph{et~al.}
\newblock \bibinfo{title}{{The 105-Month Swift-BAT All-sky Hard X-Ray Survey}}.
\newblock \emph{\bibinfo{journal}{\apjs}} \textbf{\bibinfo{volume}{235}},
  \bibinfo{pages}{4} (\bibinfo{year}{2018}).
\newblock \eprint{1801.01882}.

\bibitem{Fermi2017}
\bibinfo{author}{{Ajello}, M.} \emph{et~al.}
\newblock \bibinfo{title}{{3FHL: The Third Catalog of Hard Fermi-LAT Sources}}.
\newblock \emph{\bibinfo{journal}{\apjs}} \textbf{\bibinfo{volume}{232}},
  \bibinfo{pages}{18} (\bibinfo{year}{2017}).
\newblock \eprint{1702.00664}.

\bibitem{Bourriche2023}
\bibinfo{author}{Bourriche, N.} \& \bibinfo{author}{Capel, F.}
\newblock \bibinfo{title}{{Cosmic cartography with UHECRs: Source constraints
  from individual events at the highest energies}}.
\newblock \emph{\bibinfo{journal}{PoS}} \textbf{\bibinfo{volume}{ICRC2023}},
  \bibinfo{pages}{362} (\bibinfo{year}{2023}).

\end{thebibliography}

\clearpage
\begin{addendum}
\item [Acknowledgements]
We thank the useful discussions with Ruoyu Liu, Tianqi Huang, Bing Theodore Zhang, Gwenael Giacinti, Jun Xu, Xuening Bai, Ruizhi Yang and Xiangyu Wang.
We thank the useful comments from the anonymous referees.
H.N.H. is supported
by Project for Young Scientists in Basic Research of
Chinese Academy of Sciences (No. YSBR-061), and
by NSFC under the grants No. 12173091, No.12321003 and No.12333006, and the Strategic Priority Research Program of the Chinese Academy of Sciences No.XDB0550400.
E.K. is funded by Chinese Academy of Sciences President’s International Fellowship Initiative. Grant No. 2025PVC0021.
S.N. is supported by the ASPIRE project for top scientists, JST 'RIKEN-Berkeley mathematical quantum science initiative.'
Z.Q.X. and X.L. are supported by the Youth Innovation Promotion Association CAS.

\item [Author contributions] 
H. N. H. initiated and led the project, conducted the primary data analysis
and simulations, and wrote the majority of the manuscript. H.N. H., E. K. and K. K. D.work together to refine the methodology. Y. Y. created Figure S3 and conducted searches for galaxies with lobes from the nearby galaxy catalog. All authors were actively involved in the discussions and approved
the scientific results.

\item [Data Availability] 
The PAO {\it phase I} data used in this paper are publicly available at https://doi.org/10.5281/zenodo.6504276. 
\item [Code Availability]
The CRpropa Software we used for the simulation of a UHECR multiplet is available at https://crpropa.desy.de.
\end{addendum}

\end{spacing}
\end{document}